\documentclass[manuscript]{aastex}

\usepackage{pifont} 
\shorttitle{Fluctuation of near-infrared background}
\shortauthors{Matsumoto et al.}

\begin{document}

\title{AKARI observation of the fluctuation of the near-infrared background.}

\author{T. Matsumoto\altaffilmark{1,2}, H. J. Seo\altaffilmark{1},  W. -S. Jeong\altaffilmark{3}, H. M. Lee\altaffilmark{1},  S. Matsuura\altaffilmark{2}, H. Matsuhara\altaffilmark{2}, S. Oyabu\altaffilmark{2,4}, J. Pyo\altaffilmark{3}, and T. Wada\altaffilmark{2}}

\altaffiltext{1}{Department of Physics and Astronomy, Seoul National University, Seoul 151-742, Korea}
\altaffiltext{2}{Department of Infrared Astrophysics, Institute of Space and Astronautical Science (ISAS), Japan Aerospace Exploration Agency (JAXA), Sagamihara, Kanagawa 252-5210, Japan}
\altaffiltext{3}{Korea Astronomy and Space Science Institute (KASI), Daejeon 305-348, Korea}
\altaffiltext{4}{present address: Department of Physics, Nagoya University, Nagoya 464-8601, Japan} 
\email{matsumo@ir.isas.jaxa.jp}

\begin{abstract}
We report a search for fluctuations of the sky brightness toward the north ecliptic pole (NEP) with the
Japanese infrared astronomical satellite AKARI, at 2.4, 3.2, and 4.1 $\mu$m. We obtained circular 
maps with $10\arcmin$ diameter field of view which clearly show a spatial structure on scale of a few hundred arcsec. 
A power spectrum analysis shows that there is a significant excess fluctuation at angular scales 
larger than $100 \arcsec$ that can't be explained by zodiacal light, diffuse Galactic light, shot noise 
of faint galaxies or clustering of low redshift galaxies. These results are consistent with observations  
at 3.6 and 4.5 $\mu$m by NASA's Spitzer Space Telescope.  The fluctuating component observed at large 
angular scales has a blue stellar spectrum which is similar to that of the spectrum of the excess 
isotropic emission observed with IRTS.  A significant spatial correlation between wavelength bands 
was found, and the slopes of the linear correlations is consistent with the spectrum of the excess fluctuation. 
These findings indicate that the detected fluctuation could be attributed to the first stars of the universe, 
i.e. pop. III stars. The observed fluctuation provides an important constraints on the era of the first stars.
\end{abstract}


\section{Introduction}  

Observations of the near-infrared background are believed to be very important for investigating the 
formation and evolution of the first stars \citep{Santos02}. A search for the cosmic near-infrared background has been 
commenced using data from the Diffuse Infrared Background Experiment (DIRBE)/Cosmic Background 
Explorer (COBE) and Near-infrared Spectrometer (NIRS)/Infrared Telescope in Space (IRTS) 
\citep{Hauser98, Cam01, Arendt03, Matsumoto05, Levenson07}. The results have consistently shown that 
there remains an isotropic emission that can hardly be explained by known sources, as long as the same
model is applied for zodiacal light. One possible origin of this excess emission is population III (pop. III) stars, 
the first stars of the universe that caused its reionization \citep{Salvaterra03, Dwek05b}. The observed excess 
emission, however, is much brighter than expected, and requires a extremely high rate of star formation during the 
pop. III era \citep{Madau05}. Furthermore, TeV-$\gamma$ observations of distant blazars favor a low level of
near-infrared extragalactic background light \citep{Dwek05a, Aharonian06, Raue09}. It has been pointed out that  
uncertainties in the modelling of zodiacal light may explain the ambiguity in the result, since zodiacal light is the 
brightest foreground component in the near-infrared range \citep{Dwek05b}. In order to obtain firm results 
on the cosmic near-infrared background, some researchers looked for fluctuations \citep{Kashlinsky02, Matsumoto05}, 
as zodiacal light is very smooth. Furthermore, fluctuation measurements are important for understanding 
the structure formation in the pop. III era.   Recently, \citet{Kashlinsky05a}  reported 
significant fluctuations at angular scales of $100\arcsec \sim 300\arcsec$ based on Spitzer data and 
concluded that they were caused by pop. III stars. However, a few studies have claimed that the 
origin of this fluctuation was the clustering of low redshift galaxies \citep{Cooray07, Chary08}. \citet{Thompson07a} 
detected a significant fluctuation at 1.1 $\mu$m and 1.6 $\mu$m with the Near-Infrared Camera and
 Multi-Object Spectrometer (NICMOS)/Hubble Space Telescope (HST), although their observation was 
 restricted to angular scales smaller than $100 \arcsec$. However, they claimed that the observed 
 fluctuation could be due to faint galaxies at  redshift  \( z < 8\) \citep{Thompson07b}. 
 In this paper, we present new observations 
 of fluctuations in sky brightness using the Japanese infrared astronomical satellite AKARI. AKARI has a cold 
 shutter to measure dark current precisely as well as a short wavelength band at 2.4 $\mu$m, which provided 
 new observational evidence on the study of the sky fluctuation. 

This paper is organized as follows. In \S 2, we provide detailed information on the observation and data 
reduction procedure. In \S3,  we describe the fluctuation analysis of the near-infrared sky, and we discuss 
the possible origins of the fluctuation in \S 4. Discussions on the implication of our data in view 
of the first stars are given in \S 5. The final section summaries our results.  

\section{Observations and data reduction}

AKARI is the first Japanese infrared astronomy satellite with a cooled 68 cm aperture telescope, launched 
into a Sun-synchronous orbit on February 22, 2006 \citep{Murakami07}. The Infrared Camera (IRC) is one of the
instruments onboard AKARI  \citep{Onaka07}, and has a frame of $10\arcmin \times 10\arcmin$ with a pixel scale of 
$1.5\arcsec$. The IRC performed a large area survey around the north ecliptic pole (NEP) \citep{Lee09}, 
and monitored a fixed position in the sky near the NEP to evaluate its performance \citep{Wada07}. The 
observed region of the sky was named NEP monitor field ($\alpha_{2000}$=17h55m24s, $\delta_{2000}$=66:37:32). 
Pointed observations were carried out twice a month, and three sets of imaging observations were performed 
for all wavelength bands of the IRC during one pointed observation.  Dithering of $\sim 10\arcsec$ was 
applied between these three sets of imaging observations. Since maneuvering of the satellite sometimes commenced 
before completing the third set of imaging observations, the last image at 4.1 $\mu$m was usually useless. 
The integration time for one image was 44.41 seconds, and the data were retrieved with Fowler 4 samplings. 
Systematic errors due to uncertainty in the the absolute calibration were 2.8, 2.5, and 3.4 \%, which were small 
compared to other errors. The details of observation mode and calibration are shown in the IRC 
User's Manual \citep{Lorente08}.

In order to detect the fluctuations in sky brightness, we used near-infrared images taken in 14 pointed observations 
between September 2006 and March 2007, during which no radiation from the earth entered the baffle and 
stray light was negligible. Incidence of earthshine into the telescope tube was detected from the temperature 
sensor on the baffle and confirmed by the IRC simultaneously as a spurious stray radiation.
Furthermore, we did not use images when the attitude control was less stable than usual, as well as those images 
suffered from a large number of  
cosmic ray hits. After all, we chose 40, 39, and 24 images from 2.4, 3.2, and 4.1 $\mu$m bands, respectively, for the analysis, 
(hereafter we list data at these three bands in order of wavelength). Total integration times were 
29.6, 28.9, and 17.8 minutes for each band, which are rather shorter than Spitzer observations  
\citep{Kashlinsky05a}.  IRC/AKARI, however, has a larger field of view, $10\arcmin$ than IRAC/Spitzer, $5.2\arcmin$.
Since pixel throughputs are almost same,  IRAC/Spitzer needs
$\sim 4$ times more observing time than IRC/AKARI  to observe the same area of the sky with the same S/N ratio.

We analyzed the sky image following the IRC User's Manual \citep{Lorente08} with some improvements. 
We estimated the dark current of the image field from that of the masked region on the array by applying the 
new method developed by \citet{Tsumura11}. 
This method provides a more reliable estimate of the dark current for those frames suffering from an after effect 
due to strong cosmic ray hits occurring  in the South Atlantic Anomaly region. We also constructed better flat fields using 
all images obtained during the period as used for the observation of the NEP 
monitor field. In total, 774, 771, and 516 images were stacked, and significantly  better flat fielding was
achieved compared to the ordinary pipeline. 
For the correction of the aspect ratio, we used the "nearest" option as interpolation type in the MAGNIFY 
task of the Image Reduction and Analysis Facility (IRAF)  \citep{Tody86} to avoid a stripe-like artifact. 

Finally we stacked the images and obtained the maps of the NEP monitor field.  Due to constraints on
attitude control, the position angles of the images were rotated by $1^{\circ}$ per a day, and by $200^{\circ}$ 
during the observation period, resulting in circular images with $10\arcmin$  diameters. This improves the quality 
of the stacked images compared to usual stacking, since flat field errors, fluctuations in sensitivity between 
pixels, residuals in dark frame subtraction, fluctuation of the zodiacal light (if any), etc. are reduced, except for the 
central part of the image and any axially symmetric structure. We also obtained dark images  with the cold shutter 
closed.  A dark image corresponding to each sky image was retrieved before or after the pointed observation.
 As a result, we obtained equal  numbers of sky and dark images for each band. For the dark images, 
 we applied  the same procedure as for the sky images and stacked them to obtain dark maps for the three 
 wavelength bands. 

In order to obtain the fluctuations in sky brightness, we masked pixels whose signal exceeded 2 $\sigma$ level. 
We repeated this 10 times after which additional clipping did not show significant changes. However, 
there remained bright pixels around the masked region due to the extended point spread function (PSF). 
In order to subtract this effect, we searched for foreground sources with a detection threshold of 2 $\sigma$ in the unmasked 
stacked images, and removed their contributions using the DAOPHOT package in the IRAF.
 At first we carefully modeled  the PSF based on the beam profiles of the bright point sources in the 
 unmasked stacked images. Using the clearly identified point sources (sources well fitted by the PSF), we estimated 
 the contribution of sources in the outer region using the model PSF and subtracted them from the sky images. 
 We were left with some point sources (sources rejected by the PSF fit) that can not be fitted with the PSF due to faintness, 
 overlapping of sources or other reasons. For these sources, we simply masked 8 neighboring pixels around their centers. 
 We identified extended objects using ground-based data for the NEP region \citep{Imai07} which have a better 
 spatial resolution. We created convolved images using AKARI's PSF and subtracted them from the sky images. 
 For these PSF subtracted images, we masked the same region in the 2 $\sigma$ clipping image. In order 
 to minimize contributions from PSFs, we further masked a layer of one pixel around the masked regions in which  
 the pixel of 2 $\sigma$ brighter level exists. 
For the dark maps, we masked the same region as with the sky maps and repeated 2 $\sigma$ clipping 10 times. 
Table 1 shows some parameters used and obtained in this analysis. Percentages of remaining pixels are 
$\sim47 \%$ for all wavelength bands, and average sky brightness are 114, 73, and 105 nWm$^{-2}$sr$^{-1}$. 
As for the sky brightness, systematic errors are not taken into account in order to show the statistical significance clearly. 

Figure 1 shows the final fluctuation maps. The upper and lower panels indicate the sky and dark maps, respectively. 
The sky maps clearly show structure, while the dark maps look like random noise.  As it is well known, a major emission 
component in the near-infrared sky is zodiacal light. We have to keep in mind the fact that the contribution of 
galactic stars is negligible, but there remains integrated light of galaxies fainter than the limiting magnitudes, and 
isotropic emission.

\section{Fluctuation spectrum}

Fluctuation analysis was first performed by \citet{Kashlinsky02} and \citet{Odenwald03} using 2MASS data, 
and followed by  \citet{Kashlinsky05a, Kashlinsky07a, Arendt10} for Spitzer data and by  
\citet{Thompson07a, Thompson07b} for HST data. 
In order to examine the angular dependence of the fluctuation observed with AKARI, we performed a
power spectrum analysis for both the sky and the dark maps (Figure 1), following the works of previous authors. 
A two-dimensional Fourier transformation was computed for the fluctuation field \( \delta F({\bf x} ) \) as a function of 
the two dimensional angular wavenumber, {\bf q}, and the two dimensional coordinate vector of the image, {\bf x}.

\[ F({\bf q}) =  \int \delta F({\bf x} ) \exp ^{(-i{\bf x} \cdot {\bf q})} d^{2}x \]

The two-dimensional power spectrum and fluctuation spectrum are given by  \( P_{2}({\bf q}) \equiv | F({\bf q})|^{2} \), and 
\( [q^{2}P_{2}({\bf q})/(2\pi)]^{1/2} \). 
Figure 2 shows the amplitude maps of  the two-dimensional fluctuation spectra
in Fourier space. The maps are symmetric but
faint annular patterns appear in 2.4 and 3.2  $\mu$m  maps. These patterns correspond to small peaks in the fluctuation spectra (Figure 3) at 
$\sim 5\arcsec$ and are probably artifacts caused by the masking pattern, since the same features appear in both sky and 
dark images. However, they do not affect the conclusion of this paper.
The one dimensional fluctuation spectrum, \( [q^{2}P_{2}/(2\pi)]^{1/2} \), is obtained by taking ensemble averages in the upper 
halves of the concentric rings of Figure 2 with equal spacing in the angular wavenumber, $q$.

Figure 3 shows the fluctuation spectra.  The upper panel shows the fluctuation spectra for both the sky (filled circles) and dark 
(open triangles) maps. Statistical errors, given by standard deviations divided by the square root of the number of the data points, are plotted. 
The fluctuation spectra of the sky maps are significantly larger than those of the dark maps. 

We subtracted the fluctuation spectra of the dark maps from that of the sky maps in quadrature, and show
the result as filled circles in the lower panel of Figure 3. The solid lines indicate the fluctuation spectra due to 
the shot noise of faint galaxies that were estimated by simulation. We first obtained galaxy counts for the 
stacked images before masking and extended them to the faint end assuming a slope of 
0.23 for logarithmic counts  \citep{Maihara01}. We constructed fluctuation maps by randomly distributing 
the galaxies fainter than the limiting magnitudes, and performed the power spectrum analysis.  
The fluctuation spectra of the shot noise of faint galaxies fit well the observed power spectra 
with limiting magnitudes (AB) of 22.9, 23.2, and 23.8. 
The decline at small angle is caused by the PSF, while the absolute value depends on limiting 
magnitude and galaxy counts. Figure 3 shows 
that a significant residual fluctuation above the shot noise level remains at large angular scales. 
Table 2 shows the numerical values of fluctuation of the observed sky, fluctuation due to shot noise by faint galaxies 
and excess fluctuation at angles larger than $100\arcsec$.

We performed a subset analysis to confirm the celestial origin of the observed structure. We divided our set of 
images into two subsets corresponding to two cases. In case A, each subset is taken as a sequence alternating 
in time, while in case B, the subsets correspond to the earlier and later halves of our set.
 For both cases, we obtained two stacked images, F1 and F2, by 
applying the same procedure as the one described previously in this section. The fluctuation spectra of the difference 
between these two stacked images are shown in Figure 4.  Case A and B are shown as squares and  
asterisks, respectively, while the fluctuation spectra of the dark maps are shown by triangles. The results 
for both subsets are consistent with those of the stacked dark maps. This indicates that the observed structure is 
indeed present in the original images and is of celestial origin. 

We also examined the impact of masking on the fluctuation spectra, since $\sim$ 53 \% of all pixels are masked. 
We constructed a common mask that includes all pixels masked in any of three wavelength bands. The fraction of remaining pixels
in images with the common mask applied is $\sim$ 32 \%. We again performed the analysis for these images, and compared the 
obtained fluctuation spectra with the original ones shown in Figure 5. The fluctuation spectra of the images using the common mask
have slightly lower than the original ones, but are consistent with the original ones within the error bars. This result assures
that the masks adopted do not cause significant artifacts and that the detected fluctuation is a real structure in the sky.

The upper panel of Figure 6 shows smoothed sky maps that were obtained by taking averages within circles of $50\arcsec$ 
diameter centered at each pixel. The smoothed maps clearly show the existence of a significant extended spatial 
structure with an angular scale of a few hundred arcsec for all wavelength bands, and the overall pattern is very 
similar. This is consistent with the excess fluctuation at large angular scales shown in Figure 3. The smoothed dark maps 
in the lower panel of Figure 6 do not show any structure besides random noise, thus they reconfirm the celestial origin
of the observed large scale structure.

In order to examine the fluctuation quantitatively, we took the averages of the excess fluctuation at angular 
scales  between $100\arcsec$ and $350\arcsec$. Figure 7 shows the spectrum of the fluctuating component  (filled circles) 
with 1 $\sigma$ errors. The characteristic feature of the fluctuating component is a blue-star-like spectrum, which 
is consistent with a Rayleigh-Jeans spectrum, $\sim\lambda^{-3}$ (straight line in Figure 7). We note that the observed 
spectrum of the fluctuating component is similar to the spectrum of the isotropic excess emission observed by IRTS 
\citep{Matsumoto05}. Open squares show the Spitzer results at 3.6 $\mu$m and 4.5 $\mu$m 
\citep{Kashlinsky05a, Kashlinsky07a}, which are basically consistent with ours. We did not plot the Spitzer 
results at longer wavelength bands, since the
 data were scattered and appeared less reliable. At the wavelength bands at 1.1 $\mu$m and 1.6 $\mu$m (NICMOS/HST), 
 \citet{Thompson07b} detected fluctuation of 0.3 and 0.4 nWm$^{-2}$sr$^{-1}$, respectively, at an angular scale of 
$ 85\arcsec$. These are lower than the values extrapolated from the AKARI data, assuming the Rayleigh-Jeans 
 spectrum, $\sim\lambda^{-3}$. However, the angular scale for NICMOS/HST is too small to have detected the 
 large-scale structure shown in the smoothed sky maps (Figure 6). Therefore, we consider it is inappropriate to 
 compare the NICMOS/HST results directly with those of AKARI and Spitzer.

We performed a two-points correlation (autocorrelation) analysis,
\( C_{F}( \theta ) =  \langle \delta F({\bf x+ \theta)}   \delta F({\bf x}) \rangle \), both for the sky and 
dark maps. In order to save computation time, we reduced the angular resolution
by adopting the average values groups of 2 $\times$ 2 pixels.  
The results, $C_{2.4}( \theta ) $, $C_{3.2}( \theta ) $, and $C_{4.1}( \theta ) $ are shown in Figure 8 as 
a function of $\theta$ in units of arcsec.  The data were in groups of 4 pixels, corresponding to
an angular resolution of $6\arcsec$. Error bars indicate statistical errors, i.e. standard deviation divided by the square root of the number
of pairs.  It must be noted that the sampling at angles larger than $300\arcsec$ is not complete, that is,
central parts of images are not counted.
Figure 8 clearly shows excess low frequency components which are common for all three
wavelength bands. The low frequency components correspond 
to the excess fluctuations at large angles, as shown in Figure 3.

We made a point-to-point correlation study  between the wavelength bands using independent data in smoothed maps. 
The result is shown in Figure 9. The correlation between 3.2 and 2.4 $\mu$m is fairly strong (correlation 
coefficient $\sim 0.73$), while that between 4.1 and 2.4 $\mu$m is somewhat weak (correlation 
coefficient $\sim 0.42$). This is probably because shot noise due to faint galaxies becomes important at 4.1 
$\mu$m. Colors of the correlating component (slope of Figure 9) normalized to be 1.0 at 2.4 $\mu$m (right ordinate) are also plotted in 
Figure 7 as open circles, which is similar to that of large scale fluctuation (filled circles).

We also obtained the cross correlation function,  
\( C_{F \otimes G}( \theta ) =  \langle \delta F({\bf x+ \theta } ) \delta G({\bf x}) \rangle \) between the maps taken at different wavelengths.
$C_{2.4 \otimes 3.2}( \theta )$ and $C_{2.4 \otimes 4.1}( \theta )$ are shown in Figure 10. Degradation of angular resolution 
and binning of the data are the same as for the two-point correlation analysis. 
Error bars indicate statistical errors, i.e. standard deviation divided by the square root of the number of pairs. 
The cross correlation functions in Figure 10 are very similar to the 2.4 $\mu$m two-points correlation, $C_{2.4}( \theta ) $. 
This is consistent with a
point-to-point correlation analysis (Figure 9), and the ratios of cross correlation functions to $C_{2.4}( \theta )$ 
are similar to the slopes in Figure 9. The result of our cross correlation analysis
implies that the fluctuation patterns are basically the same for all three wavelength bands.

\section{Origin of the excess fluctuation}

Being the main component of sky brightness, zodiacal light is a candidate for the origin of the fluctuation. 
The subset analysis provides evidence that zodiacal light is not the source of the observed fluctuation. 
Particularly in case B, the interplanetary dust observed during the first half period is almost independent 
from the one seen in the latter half period, since the interplanetary dust in the column along the line of 
sight has different orbits and 
eccentricities. If the observed fluctuation were caused by zodiacal light, the fluctuation spectra of the 
differences between these two subsets should be higher  than those of the dark images. 

The detected fluctuation levels are also higher than those expected for zodiacal light. IRAS \citep{Vrtilek95}, 
COBE \citep{Kelsall98}  and ISO \citep{Abraham97}  attempted to detect the fluctuation of zodiacal emission 
on scales from arcmin to degrees. The most stringent upper limit of 0.2 \% of the sky brightness was obtained by 
\citet{Abraham97} who measured  the fluctuation of zodiacal emission at 25 $\mu$m for a $45\arcmin \times  
45\arcmin$ field with $3\arcmin$ pixel scale. Recently, \citet{Pyo11} analyzed the smoothness of the mid-infrared sky
using the same data set as the one we used in this paper. They applied the same power 
spectrum analysis for the mid infrared bands, but did not detect any excess fluctuation over the random 
shot noise caused mainly by photon noise. We, therefore, assume the lowest fluctuation level at the 18 $\mu$m 
band to be an upper limit of fluctuations of the sky at an angle larger than $150\arcsec$, that is, $ \sim 0.02 \%$ 
of the sky brightness. Although this upper limit includes fluctuation due to zodiacal emission, diffuse 
Galactic light, integrated light of galaxies and cosmic background, we simply assume that the fluctuation of the zodiacal 
emission is lower than $ \sim 0.02 \%$ of the absolute sky brightness.

Fluctuation of zodiacal light is similar to that of zodiacal emission. Interplanetary dust clouds are thought to 
consist of three components \citep{Kelsall98}. The first is a smooth cloud, which is broadly distributed in 
the solar system. The second is dust bands composed of asteroidal collisional debris \citep{Nesvorny03}. 
The third is a circumsolar ring which is composed of dust resonantly trapped into orbit near 1 AU.  
We calculated the contributions of these components to the integrated brightness using the model by \citet{Kelsall98}. 
Figure 11 shows the volume emissivity of zodiacal light at 2.2 $\mu$m (scattering part, solid line) and 
that of zodiacal emission at 12  $\mu$m (thermal part, dashed line) along the line of sight towards the NEP. 
The volume emissivity is normalized such that the total emissivity at the origin is 1.0. Contributions of 
the dust bands (bottom), the circumsolar ring (middle) and the total emissivity (top) are shown separately. 
We used the model at the epoch 2006-09-15 when the contribution by the circumsolar ring reaches its maxumum. 
Figure 11 shows that the smooth cloud is a main contributor for both zodiacal light and emission, while 
that of the dust bands is very low although they are conspicuous at low ecliptic latitudes. 
Figure 11 clearly shows that the spatial distribution of the volume emissivity is almost the same for 
scattered light (zodiacal light) and thermal emission (zodiacal emission). This is because scattering 
phase function and temperature of the dust in the column do not change much in this direction. 
Although there is a small deviation for the total emissivity at the distance of 1 AU and beyond, and 
a systematic difference between scattering and thermal part for the circumsolar ring, their contribution to 
the total surface brightness is only a few percent. We conclude that the difference of the fluctuation 
between zodiacal light and zodiacal emission is at most a few percent, even if there exists a fluctuation 
of the dust density along the line of sight. 

More quantitatively, we estimated fluctuation levels  to be expected due to zodiacal light by constraining the upper 
limit of the fluctuation of zodiacal light to 0.02 \%.  The sky brightnesses corresponding to these upper 
limits are 0.023, 0.015, and 0.021 nWm$^{-2}$sr$^{-1}$. The expected fluctuation of zodiacal light could 
be reduced further during the stacking procedure, since AKARI moved along the earth orbit and IRC 
observed different regions of the solar system during the observation period. 
The observed fluctuations at large angular scales,  0.25, 0.11, and 0.033 nWm$^{-2}$sr$^{-1}$ (see Figure 7),
are significantly larger than those expected for zodiacal light.  

The spectrum of fluctuation observed with AKARI  shows a very blue color up to 4.1 $\mu$m (Figure 7), 
while IRTS observations delineated that  zodiacal emission becomes prominent at wavelengths 
longer than 3.5 $\mu$m  \citep{Ootsubo98}. This difference
also indicates that the fluctuation observed by AKARI is not due to zodiacal light.

Based on these arguments, we conclude that the contribution of zodiacal light to the fluctuation is negligible.

The second possible origin of the fluctuation is diffuse Galactic light (DGL), i.e. scattered starlight and thermal 
emission of the interstellar dust. Since DGL is closely related to the dust column density, we examined its 
correlation with the far-infrared map. The Far-Infrared Surveyor (FIS)/AKARI detected a clear cirrus feature 
toward NEP at 90 $\mu$m  with $\sim 30\arcsec$ angular resolution
\citep{Matsuura10}. Figure 12 shows the correlation between far-infrared and near-infrared maps with 
the  angular resolution of the AKARI data degraded to $30\arcsec$. No significant correlation was found, 
indicating that DGL is not the source of the observed near-infrared fluctuation.

The third candidate is the clustering of faint galaxies. A few authors \citep{Cooray07, Chary08} have claimed 
that the fluctuation observed with Spitzer is due to the clustering of low redshift galaxies. The blue color 
observed by AKARI is not consistent with the proposal by \citet{Chary08} which states that the fluctuating component originates 
in red dwarf galaxies. \citet{Kashlinsky07b} carefully examined the correlation between the map of these 
galaxies and those observed with Spitzer and concluded that red dwarf galaxies are not
responsible for the Spitzer result.   \citet{Sullivan07} 
examined the contribution of the clustering of faint galaxies to the sky fluctuation using the catalog of galaxies. 
Based on their work, we estimated that the fluctuation at 2.4 $\mu$m caused by clustering of galaxies fainter than 
the limiting magnitude of this study at multipole, $l \sim 1,000$ ($\sim 600\arcsec$), to be  
$\sim$ 0.03 nWm$^{-2}$sr$^{-1}$. This is significantly lower than the fluctuation detected at 2.4 $\mu$m. 
As for the L band, Figure 8 in \citet{Sullivan07} shows that the fluctuation due to clustering is fairly weak compared 
with the one observed by Spitzer \citep{Kashlinsky05a, Kashlinsky07a}. These results indicate that the 
clustering of low redshift faint galaxies is not likely to be the source of the observed fluctuation.

Contribution of faint galaxies to the background is studied by several authors 
 \citep{Totani00, Totani01, Maihara01, Keenan10}. According to the recent observation by \citet{Keenan10}, 
the integrated light of galaxies in K band amounts to $10.0 \pm 0.8$ nWm$^{-2}$sr$^{-1}$. 
Galaxies of $\sim$ 18 mag are mostly responsible for the integrated brightness, and the galaxies fainter  
than that contribute even less. The integrated brightness at 2.4 $\mu$m due to galaxies fainter than the limiting 
magnitude of 22.9 mag is estimated to be at most 0.7 nWm$^{-2}$sr$^{-1}$  \citep{Keenan10}. 
Observed fluctuation at large angular scales, 0.25 nWm$^{-2}$sr$^{-1}$, is a fraction of this brightness. 
\citet{Sullivan07} estimated the fluctuation due to clustering of galaxies with K magnitudes between 
17 and 19 mag.  The fluctuation obtained was 0.1 nWm$^{-2}$sr$^{-1}$, while the integrated light of 
these galaxies is 4 nWm$^{-2}$sr$^{-1}$ \citep{Keenan10}, that is, the fluctuation is 1/40 of the integrated light 
of galaxies. Furthermore, the fainter galaxies contribute less to the clustering. Compared with 
this result, the fluctuation observed with AKARI is fairly large. The situation is more conspicuous for the 
case of NICMOS/HST observations \citep{Thompson07b}, since their limiting magnitudes, 28.5 mag, are much 
fainter than  the one of the AKARI observation.  
The integrated light in H band due to galaxies  fainter than 28.5 mag is $\sim$ 0.04 nWm$^{-2}$sr$^{-1}$  
\citep{Keenan10}, while the observed fluctuation is 0.4 nWm$^{-2}$sr$^{-1}$. 
The observed fluctuation is ten times brighter than the integrated light of galaxies, which indicates that the
origin of the observed fluctuation is not the low redshift galaxies.  
These considerations imply that a simple extrapolation of galaxy counts 
hardly explains fluctuation  observed, and that there must exist a new component of infrared sources with magnitudes 
fainter than the current limit. 

Accreting black holes could be another candidate if they are abundant 
in the early universe, since they show blue spectra. However, the bulk of the light will be in much shorter 
wavelengths like X-rays, since the accretion disks must be very hot. 

To summarize, we found that the observed near-infrared fluctuation at large angular scales can hardly be 
explained by known sources. We now regard that the most probable origin of the fluctuation is pop. III stars,
the first stars of the universe. 

\section{Discussions}
AKARI observations have shown that there existed a large scale structure even at the pop. III era. Since the 
fluctuation power is nearly constant at scales of a few hundred arcsec, 
the large scale structure at the pop. III era could be more extended. 
The co-moving distance for $10\arcmin$ at $\it z$ $\sim$10 is  $\sim$ 30 Mpc which is a scale similar to 
the local super cluster in the present universe. Theoretical studies
of sky fluctuation due to the first stars have been predicted by several authors \citep{Cooray04, Kashlinsky04, Fernandez10}.
 
\citet{Cooray04}  presented the first theoretical study on fluctuations caused by pop. III stars. They assumed 
biased star formation which traces the density distribution of the dark matter. Although the absolute level of the 
fluctuation spectrum has a wide range depending on the physical parameters, the overall shapes 
have almost the same convex feature with a flat peak at $\it l \sim$1,000 
($\sim 10\arcmin$) and a turn-over towards large angles. We regard this flat peak as the most significant excess 
fluctuation observed by AKARI  which is mainly due to the pop. III stars.
The fluctuating power observed with AKARI is marginally consistent with a pessimistic estimate (Figure 3 in \citealt{Cooray04}). 
 The wavelength dependence of fluctuation power of the model shows fairly blue color which is basically consistent with AKARI observation. 
 The fluctuation observed with AKARI could be interpreted within the framework of this model. 
 However, detection of the turn-over at angular scale larger than $10\arcmin$ is essential. 

 Recently, \citet{Fernandez10} made a numerical simulation of the cosmic near-infrared background caused by 
 early populations (pop. II and III stars), assuming a halo mass of 2  $\times$ 10$^{9}$ $M_\odot$. 
 They calculated both amplitude and fluctuation, taking into account stellar emission and emission 
 from ionized gas surrounding stars.  As for the fluctuation spectrum, they predicted a monotonic decrease 
 of the fluctuation towards large angles, which is not consistent with our result. This discrepancy requires some 
 kind of modification of the theory. However, their result provides valuable information on the physical 
 conditions of early populations which are independent from halo mass.
 
A spectrum of the sky brightness at 2 $\sim$ 4 $\mu$m obtained by \citet{Fernandez10} shows a blue color, 
and the main emission component is attributed to stars. Furthermore, the ratio of the fluctuating power to the absolute 
sky brightness, $\delta$I/I, does not depend on the wavelength. This justifies  to regard the spectral shape 
of absolute brightness to be same as  that of the fluctuation. In Figure 7, the spectrum of Figure 20 of  
\citet{Fernandez10} is plotted as a dotted line.  Their spectrum is a little redder than, but is qualitatively consistent 
with, the AKARI observation.
A good correlation for the fluctuations between 
the wavelength bands observed with AKARI supports the idea that the emission source is stellar emission from 
early populations, and that the nearest stars provide the largest contribution to the sky brightness. 
$\delta$I/I estimated by \citet{Fernandez10} is $\sim$ 0.02 at a few hundred arcsec for the largest fluctuation. 
The observed fluctuation at 2.4 $\mu$m, $\sim$ 0.25 nWm$^{-2}$sr$^{-1}$, renders the absolute sky brightness 
to be brighter than $\sim$ 13 nWm$^{-2}$sr$^{-1}$, which is consistent with the excess brightness observed 
with IRTS and COBE  \citep{Matsumoto05}.

All theoretical models \citep{Santos02, Salvaterra03, Dwek05b, Cooray04, Fernandez10} predict
a clear peak due to the redshifted  Ly$\alpha$ from the ionized gas surrounding first stars. 
The wavelength of this peak depends on the redshift of the nearest first stars. 
Since there is no signature of redshifted Ly$\alpha$ in AKARI data, it must appear below 2 $\mu$m, that is, 
the redshift of the nearest first stars is less than 15. Detection of the redshifted  Ly$\alpha$ 
both in the amplitude and the fluctuation will be a key issue to delineate the epoch of formation of the first stars. 
 
The fluctuation observed with AKARI provides an evidence for light from the first stars. However, more 
observational results are required in order to investigate the physical conditions and the star formation history of the first stars. 
In a forthcoming paper, we will present a fluctuation analysis at larger angles than this paper based on 
the AKARI NEP wide field survey. The sounding rocket experiment, Cosmic Infrared Background Experiment (CIBER), 
will provide the spectrum of the sky brightness around 1 $\mu$m and the fluctuation at $1^{\circ}$ scale at the 
0.8  $\mu$m and 1.6 $\mu$m band \citep{Bock06}.  The Korean infrared satellite, Multi Purpose Infrared 
Imaging System (MIRIS) \citep{Han10} will observe the fluctuations at even larger angular scale ($\sim$ $10^{\circ}$). 
These space observations will certainly improve our understanding of the first stars of the universe.

\section{Summary}
We performed a fluctuation analysis for  a sky region with a diameter of $10\arcmin$ towards the NEP at  2.4, 3.2, and 4.1 $\mu$m 
based on observations with AKARI. A power spectrum analysis indicates 
that there exists a significant  excess fluctuation at angular scale larger than $100\arcsec$ that can't be 
explained by zodiacal light, diffuse Galactic light, shot noise from faint galaxies or clustering of low redshift galaxies. 
The fluctuating component at large angles has a blue stellar spectrum, and shows a significant spatial correlation 
between adjacent wavelength bands. 

The fluctuation observed with AKARI and NICMOS/HST \citep{Thompson07b} is considerably higher than the one  
expected from galaxy counts. This implies that a simple extrapolation of galaxy counts hardly explains the observed 
fluctuation, and that there must be a new component of infrared sources at the faint end.

The most probable origin of excess fluctuation is pop. III stars, the first stars of the universe. The detected 
fluctuation can be explained with biased star formation which traces dark matter \citep{Cooray04}. 
A blue color of the observed fluctuation and a significant spatial correlation between wavelength bands 
imply that the source of emission is stellar emission from early populations \citep{Fernandez10}. 

The angular scale of $10\arcmin$ at $\it z$ $\sim$10 corresponds to a comoving distance of 30 Mpc, a scale similar 
to local super cluster in the present universe.

The detection of a turn-over of fluctuation spectra at large angles and the spectral peak of redshifted Ly$\alpha$ at 
wavelengths shorter than 2 $\mu$m is a key issue for future observations in order to improve the investigation of 
the pop. III star formation. 

\acknowledgments
\section*{ACKNOWLEDGMENTS}
We thank the IRC/AKARI team for their encouragement and helpful discussions. Thanks also to Professors 
A. Cooray, E. Komatsu and A. Ferrara for valuable discussions and comments. This work was supported 
by JSPS Grant-in-aid (18204018). HML was supported by NRF grant No. 2006-341-C00018.

\clearpage

\begin{table}
 \caption{Parameters used and obtained in this study}
 \begin{center}
   \begin{tabular}{lrrr}
    \hline
    \hline
    Item & 2.4$\mu$m & 3.2$\mu$m & 4.1$\mu$m \\
    \hline
    
    Number of stacked images	& 40 & 39 & 24\\
    Total integration time (minutes)	& 29.6  & 28.9 & 17.8 \\
    Percentage of remaining pixels (\%)	  & 47.7 & 46.7  & 47.1 \\
    Limiting magnitude (AB mag) 	& 22.9 & 23.2 & 23.8 \\
    Average sky brightness (nWm$^{-2}$sr$^{-1}$) \tablenotemark{a}      & 114 & 73 & 105 \\
    1 $\sigma$ fluctuation, sky (nWm$^{-2}$sr$^{-1}$) \tablenotemark{a}      & 2.91 & 1.72 & 0.96 \\ 
    1 $\sigma$ fluctuation, dark (nWm$^{-2}$sr$^{-1}$) \tablenotemark{a}      & 1.98 &  1.15 & 0.59 \\ 
    
    \hline
   \end{tabular}
   \tablenotetext{a}{Systematic errors are not taken into account.} 
 \end{center}
\end{table}

\clearpage

\clearpage

\begin{deluxetable}{ccccccccccccc}
\tabletypesize{\tiny}
\rotate
\tablecaption{Fluctuation of the observed sky, fluctuation due to faint galaxies (shot noise) and excess fluctuation in units of nWm$^{-2}$sr$^{-1}$}
\tablewidth{0pt}
\tablehead{
\colhead{}&
\colhead{}&
\colhead{2.4 $\mu$m} &
\colhead{}&
\colhead{}&
\colhead{}&
\colhead{ 3.2 $\mu$m}&
\colhead{}&
\colhead{}&
\colhead{}&
\colhead{4.1 $\mu$m} &
\colhead{}& \\
\cline{2-4}
\cline{6-8}
\cline{10-12} \\
\colhead{angle (arcsec)} & \colhead{sky fluctuation} & \colhead{shot noise} & \colhead{excess fluctuation} &&
  \colhead{sky fluctuation} & \colhead{shot noise} & \colhead{excess fluctuation} &&
  \colhead{sky fluctuation} & \colhead{shot noise} & \colhead{excess fluctuation}
 }
\startdata
111 & 0.35 $\pm$ 0.082 & 0.13 $\pm$ 0.0026 & 0.33 $\pm$ 0.089 & & 0.16 $\pm$ 0.035 & 0.08 $\pm$ 0.0016 & 0.14 $\pm$.041 & &0.078$\pm$0.02 & 0.045$\pm$0.0009 & 0.063$\pm$0.025 \\
125 & 0.33 $\pm$ 0.043 & 0.12 $\pm$ 0.0017 & 0.31 $\pm$ 0.046 & & 0.16 $\pm$ 0.041 & 0.075 $\pm$ 0.0012 & 0.14 $\pm$ 0.047 & & 0.048 $\pm$ 0.011 & 0.039 $\pm$ 0.0006 & 0.027 $\pm$ 0.019 \\
143 & 0.26 $\pm$ 0.035 & 0.10 $\pm$ 0.0015 & 0.24 $\pm$ 0.038 & & 0.13 $\pm$ 0.017 & 0.064 $\pm$ 0.0010 & 0.11 $\pm$ 0.020 & & 0.040 $\pm$ 0.0075 & 0.034 $\pm$ 0.0005 & 0.021 $\pm$ 0.014 \\
167 & 0.20 $\pm$ 0.050 & 0.087 $\pm$ 0.0021 & 0.18 $\pm$ 0.056 & & 0.079 $\pm$ 0.034 & 0.055 $\pm$ 0.0014 & 0.057 $\pm$ 0.047 & & 0.039 $\pm$ 0.019 & 0.030 $\pm$ 0.0007 & 0.025 $\pm$ 0.029 \\
200 & 0.093 $\pm$ 0.023 & 0.070 $\pm$ 0.0012 & 0.061$\pm$ 0.035 & & 0.082 $\pm$ 0.016 & 0.044 $\pm$ 0.0007 & 0.068 $\pm$ 0.019 & & 0.031 $\pm$ 0.0070 & 0.025 $\pm$ 0.0005 & 0.019 $\pm$ 0.011 \\
250 & 0.29 $\pm$ 0.045 & 0.058 $\pm$ 0.0015 & 0.29 $\pm$ 0.046 & & 0.13 $\pm$ 0.029 & 0.036 $\pm$ 0.0009 & 0.12 $\pm$ 0.030 & & 0.047 $\pm$ 0.010 & 0.020 $\pm$ 0.0005 & 0.042 $\pm$ 0.011 \\
333 & 0.29 $\pm$ 0.12 & 0.041 $\pm$ 0.0013 & 0.29 $\pm$ 0.12 & & 0.13 $\pm$ 0.046 & 0.027 $\pm$ 0.0009 & 0.13 $\pm$ 0.047 & & 0.029 $\pm$ 0.012 & 0.016 $\pm$ 0.0005 & 0.025 $\pm$ 0.015 \\
\enddata
\tablecomments{Errors indicate 1 $\sigma$ error, i.e. standard deviation divided by square root of the number of data points for each ensemble.}
\end{deluxetable}

\clearpage

\begin{figure}
\epsscale{1.0}
\plotone{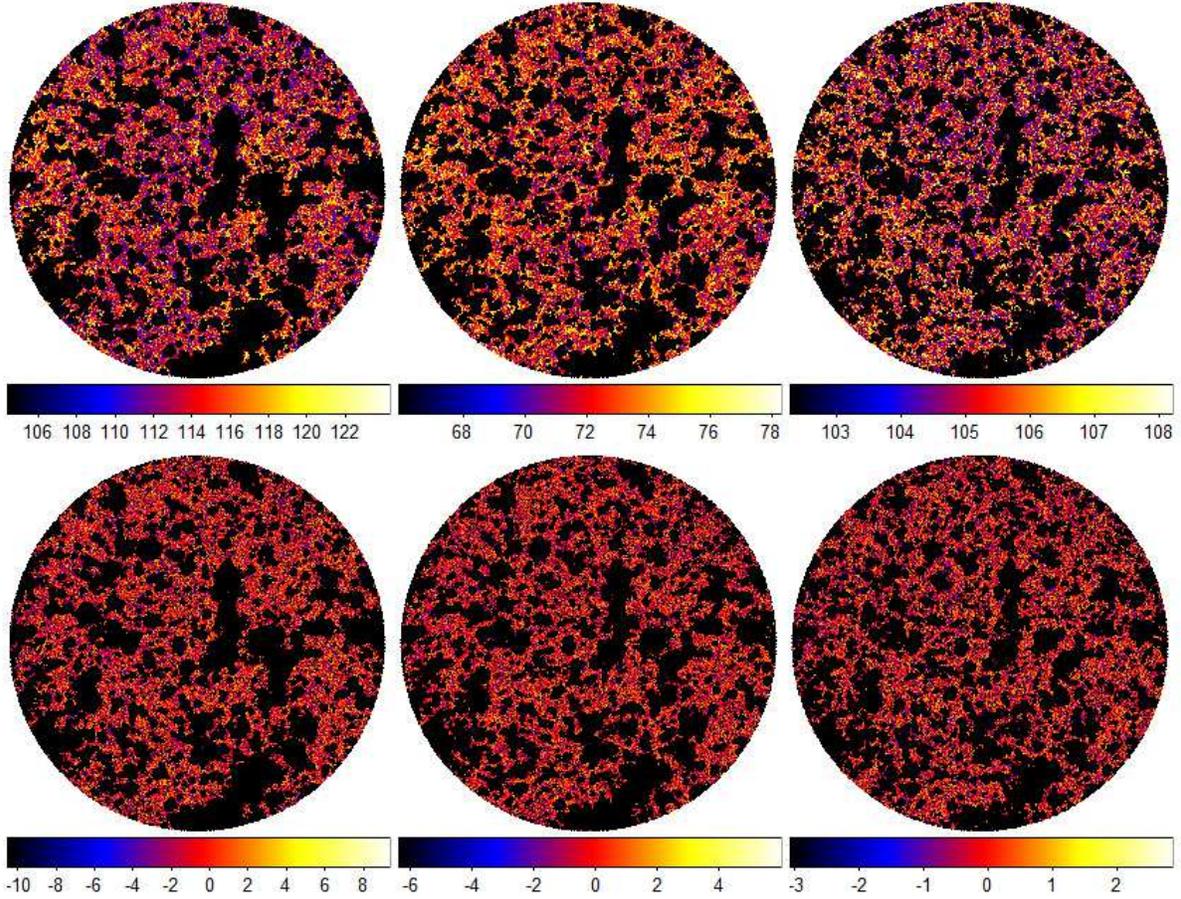}
\caption{Fluctuation maps obtained by stacking. The diameter of the circles is $600\arcsec$. Maps 
correspond to 2.4, 3.2, and 4.1 $\mu$m from the left to the right. Upper panels indicate sky maps, while 
lower panels show dark maps.  The unit is nWm$^{-2}$sr$^{-1}$.  The color scales shown in the 
bar below each map are chosen so that sky maps and dark maps cover the same ranges of sky brightness.  \label{fig1}} 
\end{figure}

\clearpage

\begin{figure}
\epsscale{1.0}
\plotone{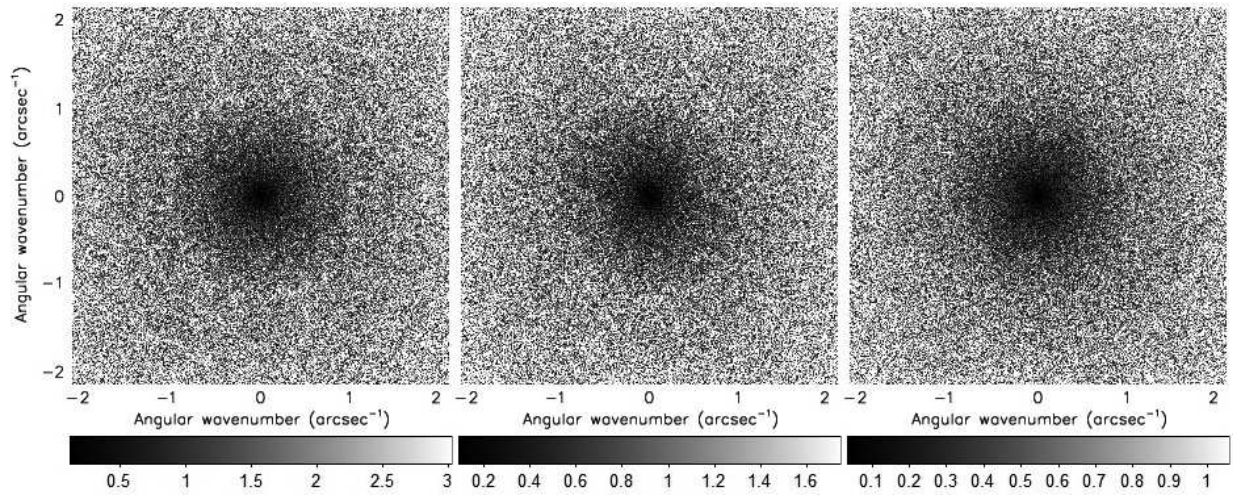}
\caption{Amplitude maps of the two-dimensional fluctuation spectra, \( [q^{2}P_{2}({\bf q})/(2\pi)]^{1/2} \), in Fourier space.
 The results for the 2.4, 3.2, and 4.1 $\mu$m band are shown from the left to the right. 
 The gray scales bars below each map indicate amplitudes of the fluctuation in units of nWm$^{-2}$sr$^{-1}$. \label{fig2}} 
\end{figure}

\clearpage

\begin{figure}
\epsscale{1.0}
\plotone{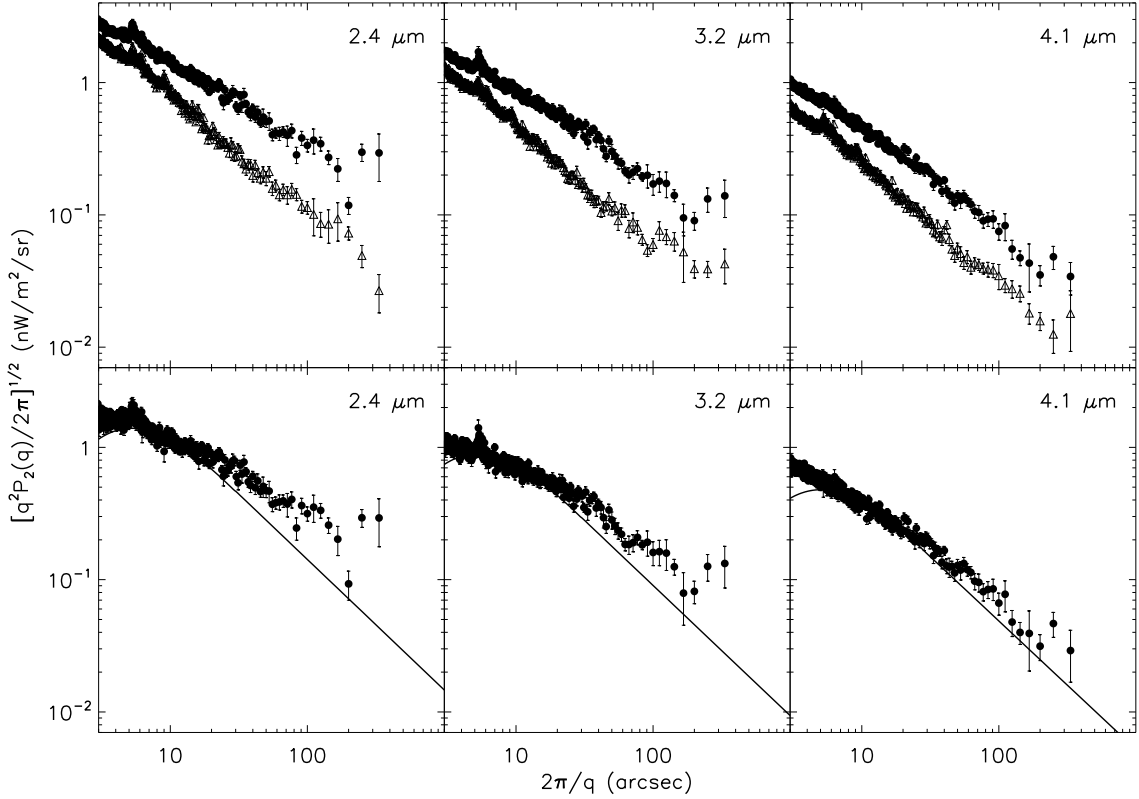}
\caption{The upper panel shows the one dimensional fluctuation spectra, \( [q^{2}P_{2}/(2\pi)]^{1/2} \) in unit of nWm$^{-2}$sr$^{-1}$, 
obtained by two-dimensional Fourier analysis as a function of angular scale \( (2\pi)/q\). Graphs correspond to the 2.4, 3.2, and 4.1 
$\mu$m band from the left to the right. Filled circles and open triangles show the fluctuation spectra for sky and 
dark maps, respectively. The lower panel shows the fluctuation spectra of the sky after subtracting those of dark maps 
in quadrature. The straight lines indicate the fluctuation spectra of shot noise due to unresolved faint galaxies.  
All error bars represent 1 $\sigma$ error. \label{fig3}} 
\end{figure}

\clearpage

\begin{figure}
\epsscale{1.0}
\plotone{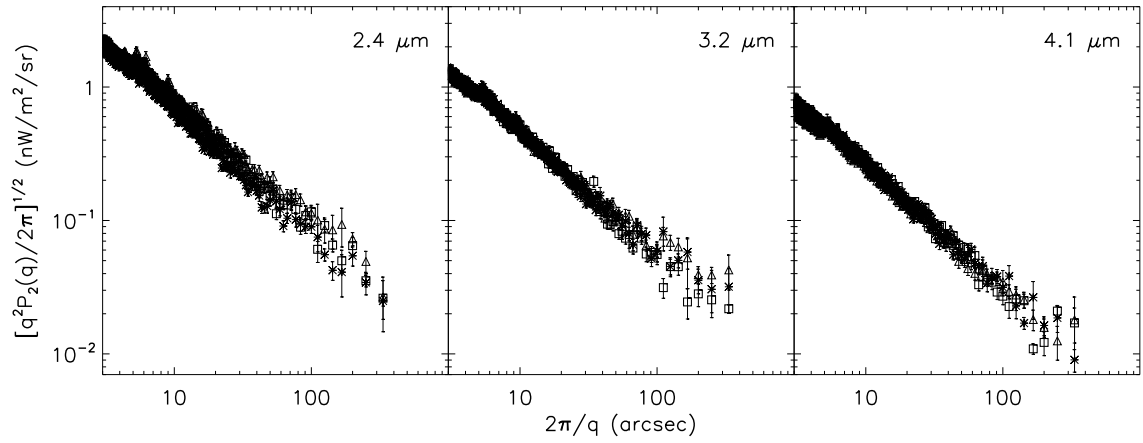}
\caption{Results of subset analysis. The squares indicate the fluctuation spectra for the subtracted image for the  
case A subset, while the asterisks for the case B subset. The triangles  show the fluctuation spectra for the dark maps. \label{fig4}}
\end{figure}

\clearpage

\begin{figure}
\epsscale{1.0}
\plotone{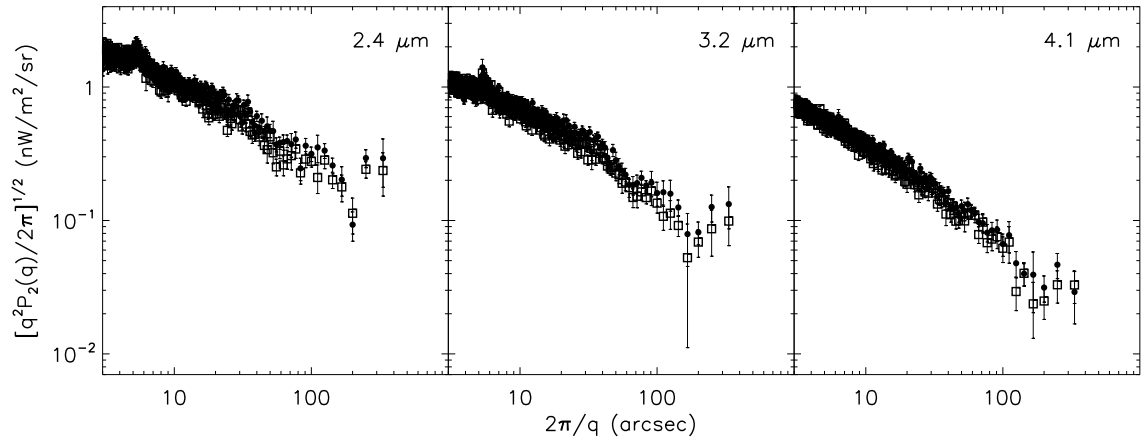}
\caption{Original fluctuation spectra (filled circles) are compared with those using common mask (open squares). 
The results for the 2.4, 3.2, and 4.1 $\mu$m band are shown from the left to the right. \label{fig5}}
\end{figure}

\clearpage

\begin{figure}
\epsscale{1.0}
\plotone{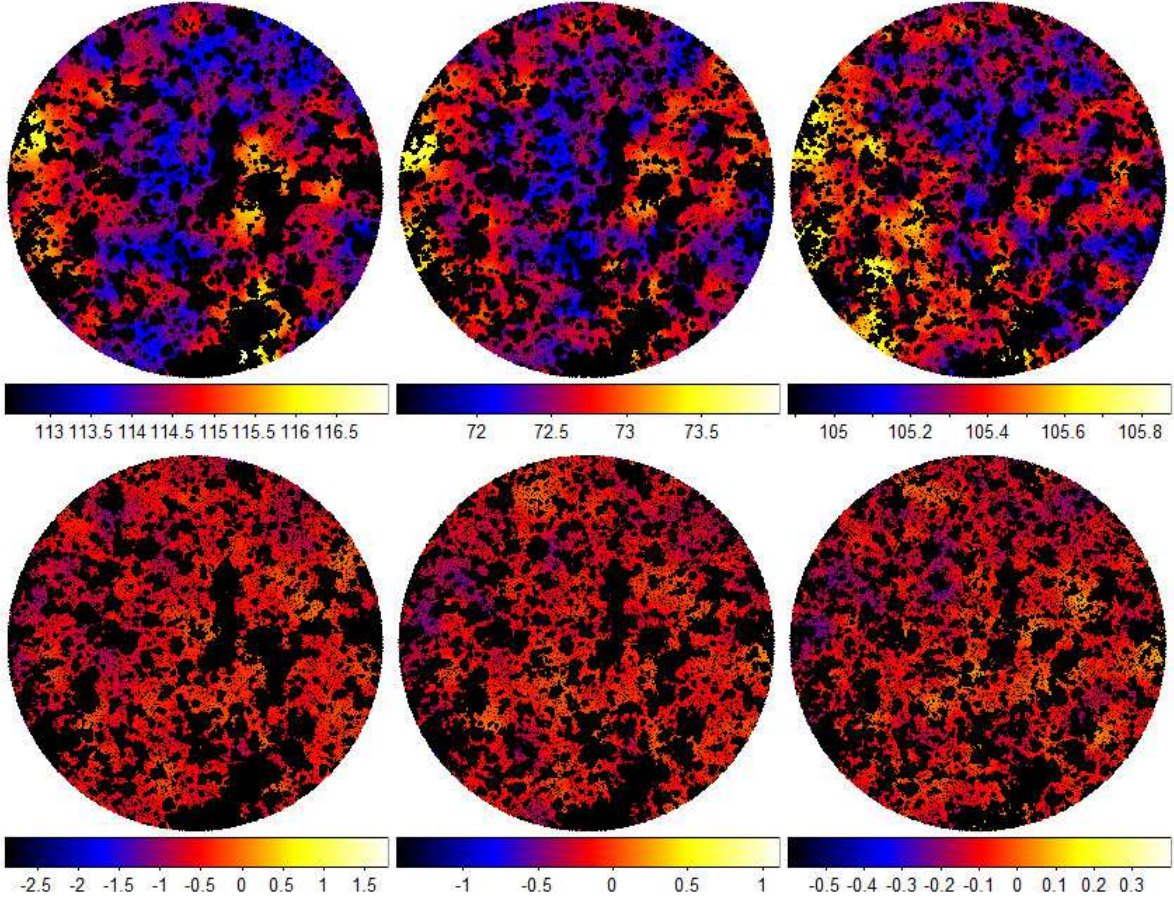}
\caption{The upper panel shows smoothed sky maps obtained by averaging pixels within a $50\arcsec$ diameter 
circle centered on each pixel. The lower panel shows smoothed dark maps obtained by the same 
procedure for the dark maps. The color scales shown in the bar below each map are chosen such that sky 
maps and dark maps have the same range of sky brightness.\label{fig6}}
\end{figure}

\clearpage

\begin{figure}
\epsscale{0.8}
\plotone{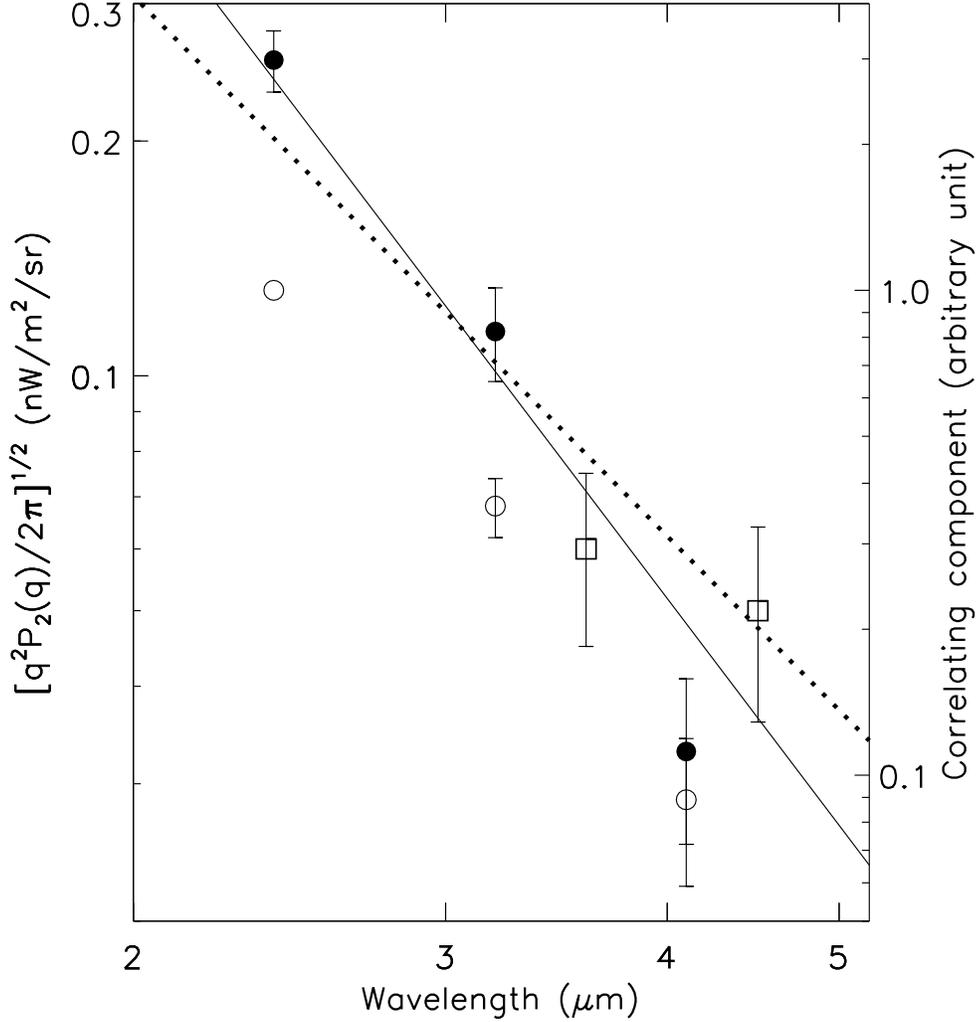}
\caption{The spectrum of the average fluctuation at large angles ($100\arcsec$ $\sim$ $350\arcsec$) (shown by filled circles) is 
compared with the Spitzer results (open squares). The open circles represent the spectrum of the correlating 
component normalized to the 2.4 $\mu$m band. The solid line indicates a Rayleigh-Jeans spectrum ($\sim\lambda^{-3}$), 
while the dotted line indicates the spectrum in Figure 20 of \citet{Fernandez10}. The vertical bars show 1 $\sigma$ error. \label{fig7}}
\end{figure}

\clearpage

\begin{figure}
\epsscale{0.8}
\plotone{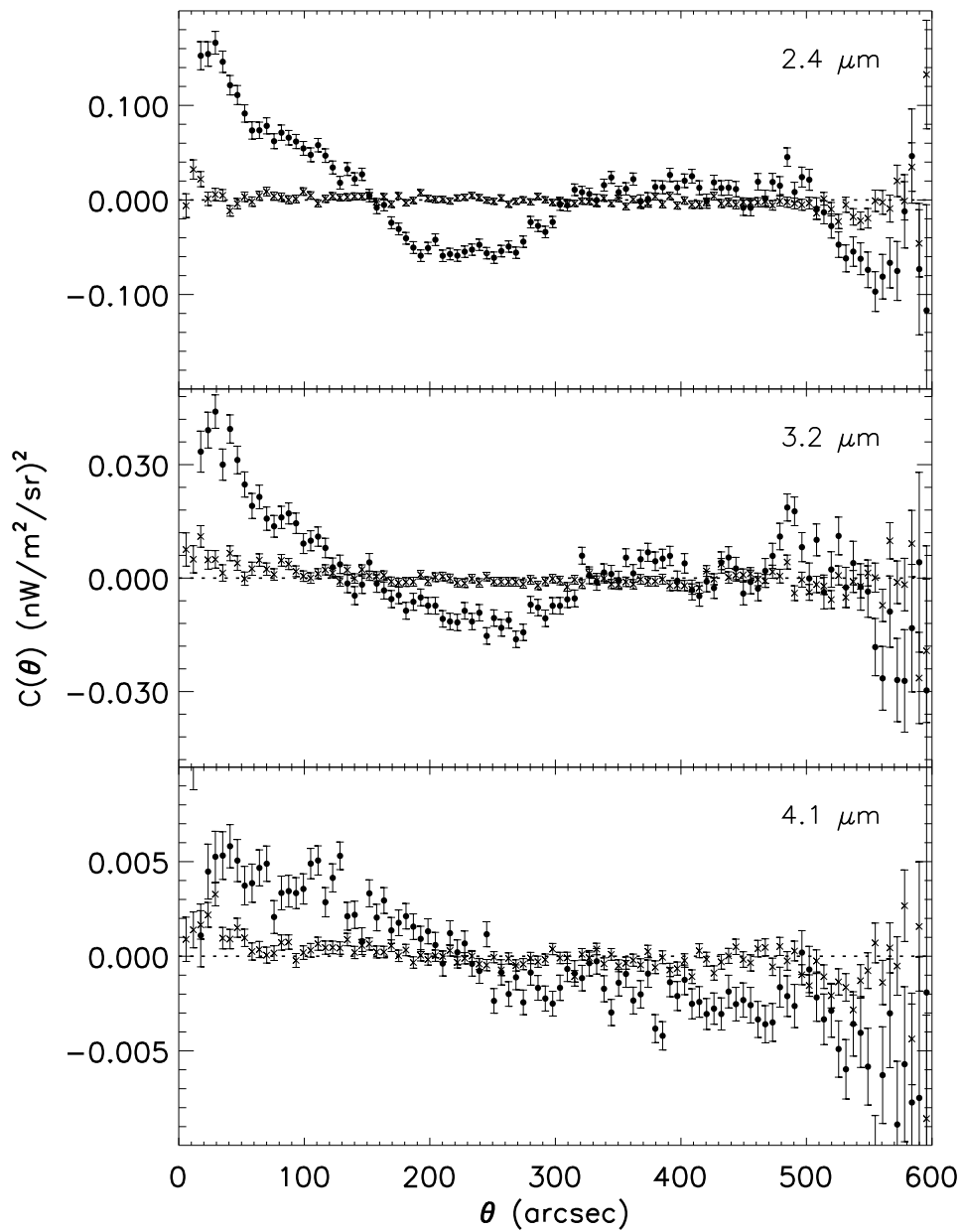}
\caption{Two-point correlation functions, $C_{2.4}( \theta ) $, $C_{3.2}( \theta ) $, and $C_{4.1}( \theta )$,
are shown from the top to the bottom. Filled circles and crosses correspond to the two-point correlation 
functions for sky and dark maps, respectively. Error bars indicate statistical errors, i.e. standard deviation divided by square root of number
of pairs. \label{fig8}}   
\end{figure}

\clearpage

\begin{figure}
\epsscale{0.8}
\plotone{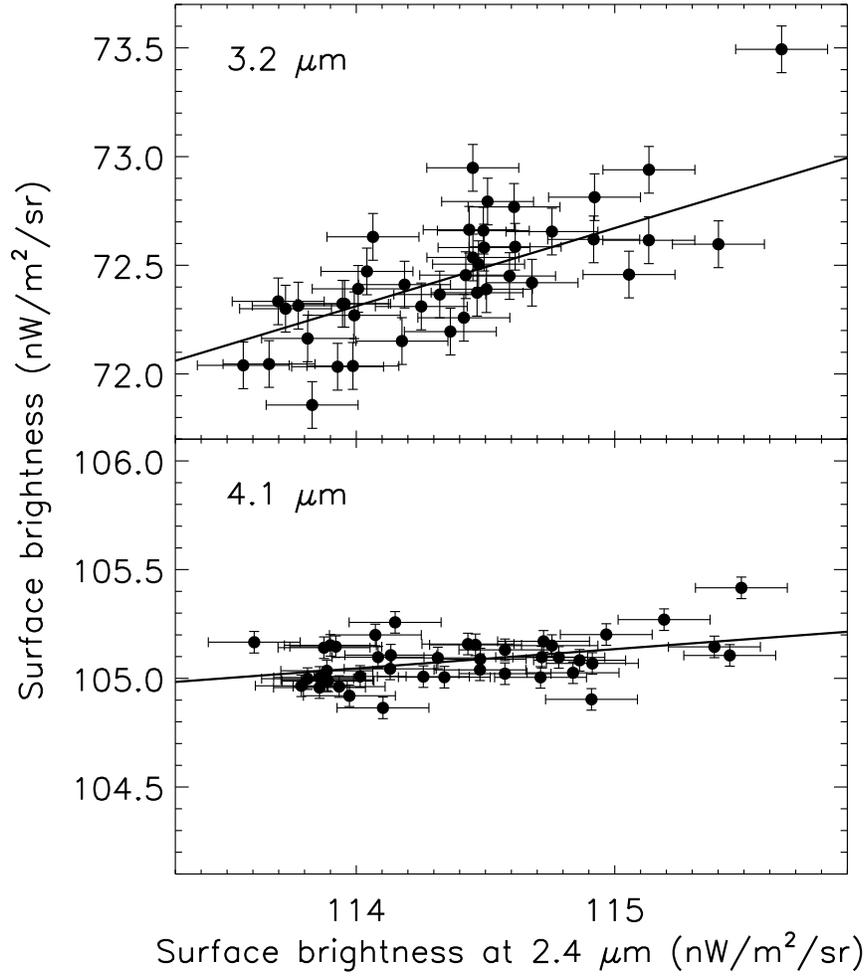}
\caption{Point-to point correlation between wavelength bands for smoothed maps. The upper panel shows the correlation 
diagram for  the 3.2 and 2.4 $\mu$m bands, while the lower one shows that for the 4.1 and 2.4 $\mu$m bands.  
Only independent data points are plotted. Error bars represent the standard deviations of the smoothed dark maps. 
The straight lines show the results of linear fits. \label{fig9}}
\end{figure}

\clearpage

\begin{figure}
\epsscale{1.0}
\plotone{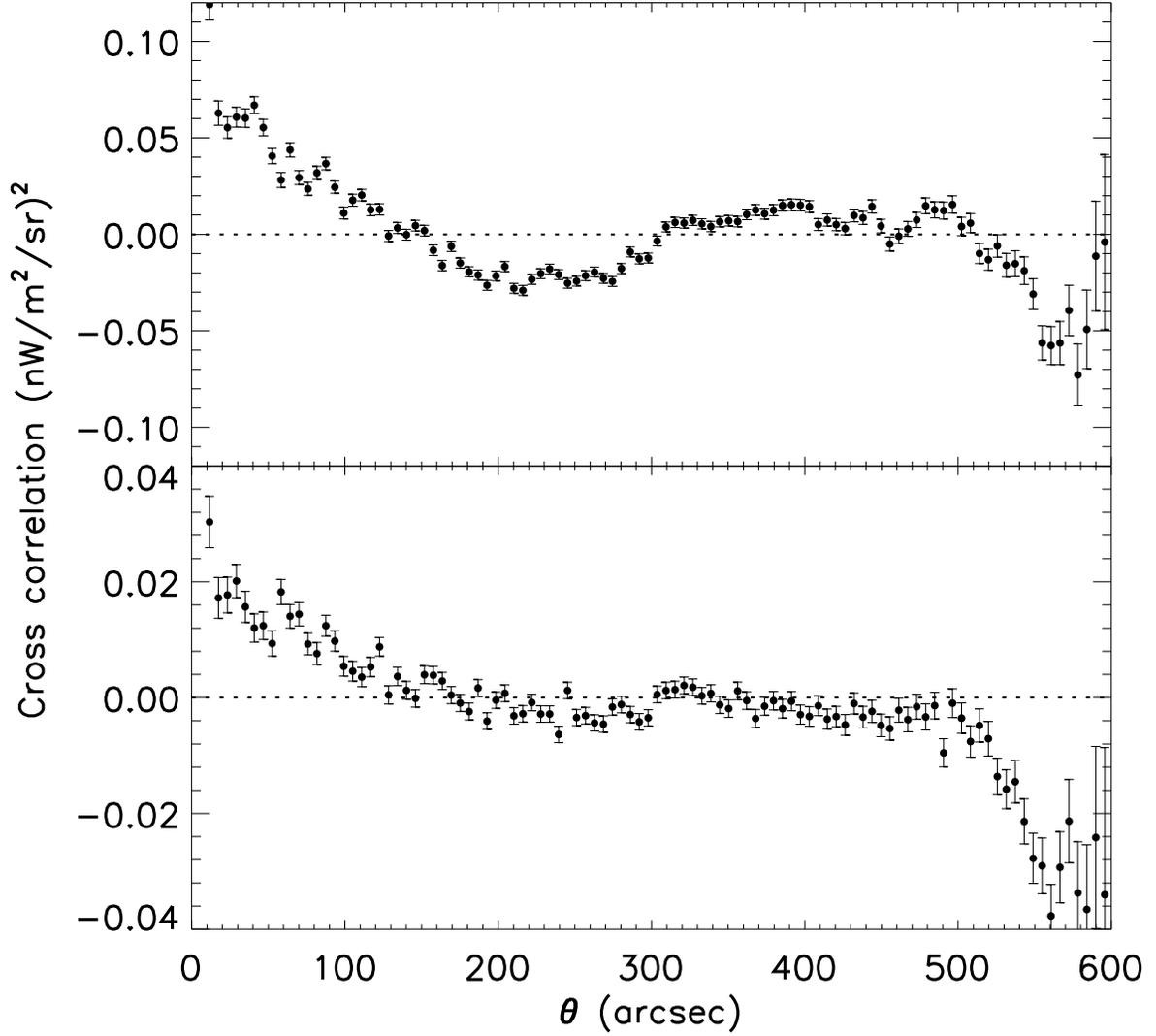}
\caption{Cross correlation functions, $C_{2.4 \otimes 3.2}( \theta )$ (upper panel) and $C_{2.4 \otimes 4.1}( \theta )$ 
(lower panel). Error bars indicate statistical errors, i.e. standard deviation divided by square root of number
of pairs.\label{fig10}}
\end{figure}

\clearpage

\begin{figure}
\epsscale{1.0}
\plotone{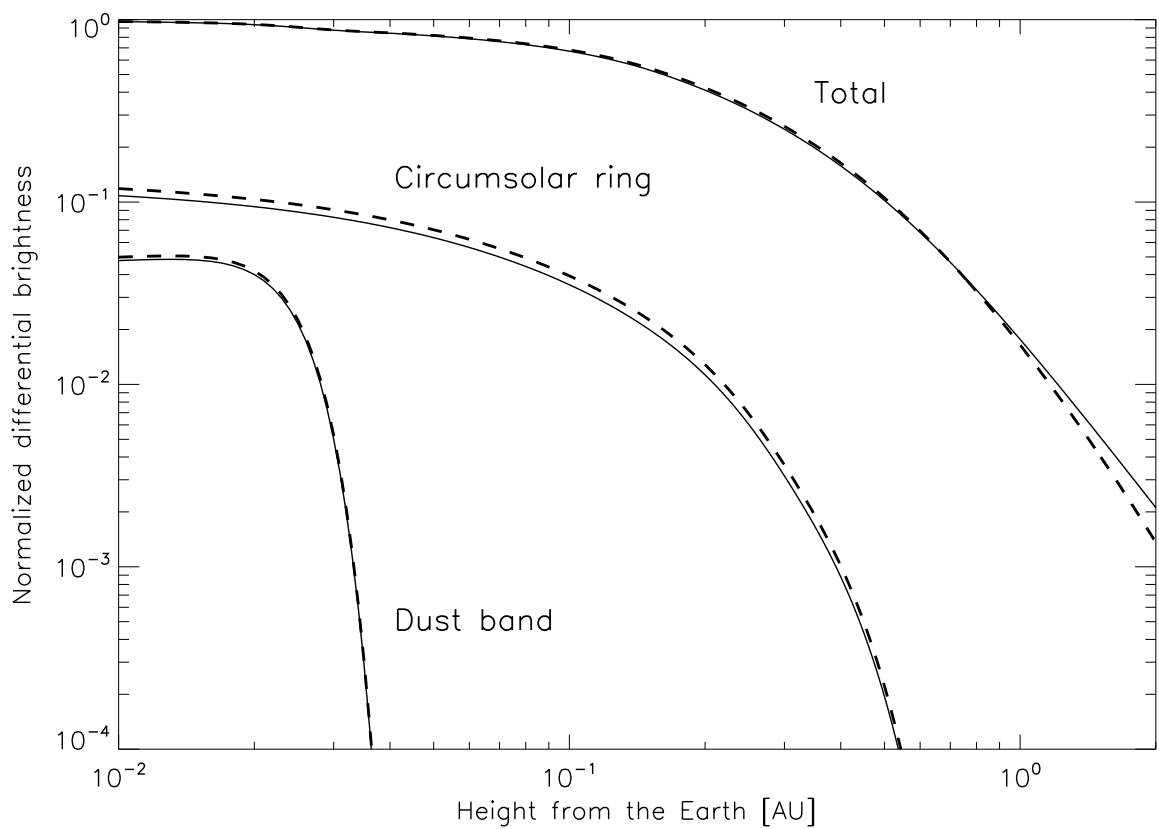}
\caption{The volume emissivity of zodiacal light at 2.2 $\mu$m (solid line) and zodiacal emission at 12 $\mu$m 
(dashed line) along the line of sight towards the NEP.  The vertical axis is normalized such that the total volume 
emissivity at the origin is 1.0. The top curve shows the total emissivity, that is, sum of the smooth cloud, the dust bands, 
and the circumsolar ring. The middle and bottom curves indicate the volume emissivities of the circumsolar ring 
and the dust bands.\label{fig11}}
\end{figure}

\clearpage

\begin{figure}
\epsscale{0.8}
\plotone{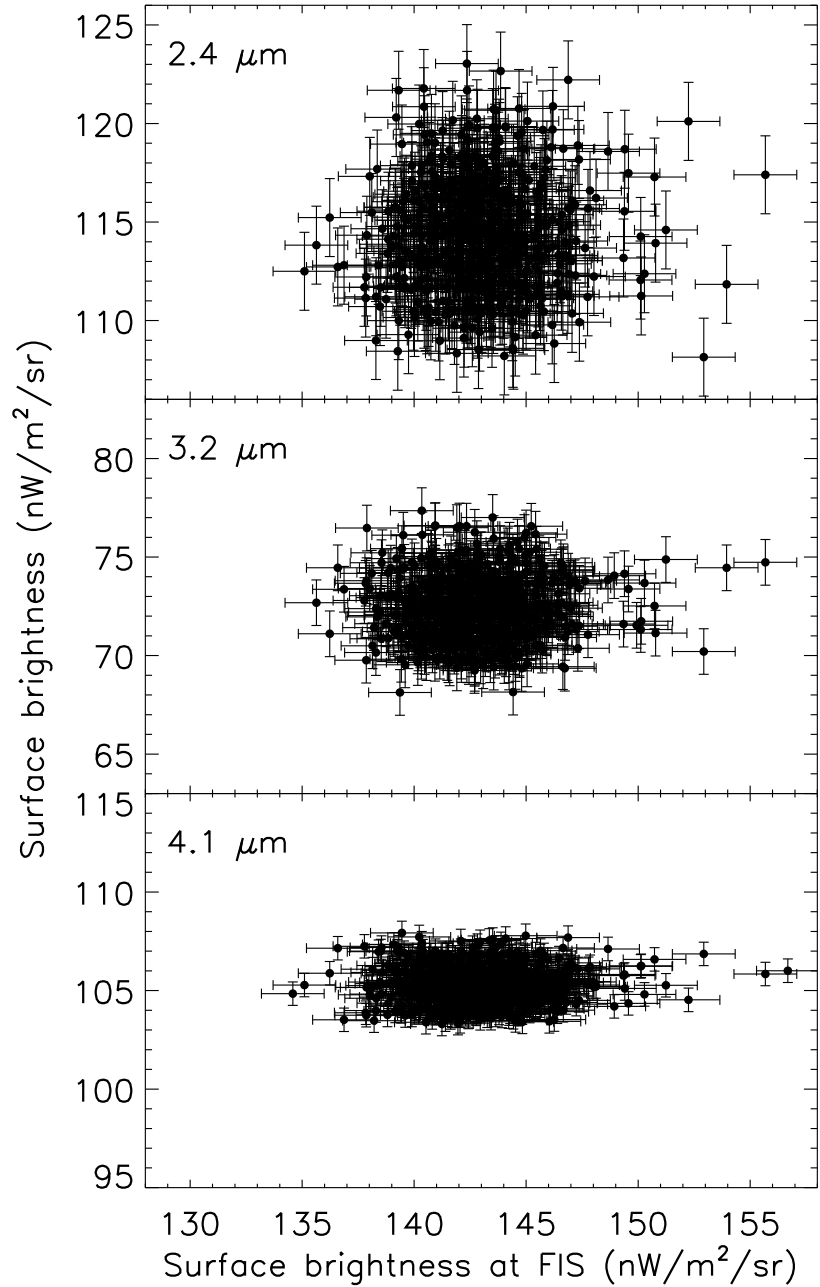}
\caption{Diagrams of correlations between AKARI maps and the far-infrared map (90 $\mu$m) observed with FIS/AKARI. 
The angular resolutions of the AKARI maps were degraded to that of the FIR observations, s$\sim 30\arcsec$. 
Error bars indicate the standard deviations of the dark maps for both observations.\label{fig12}}
\end{figure}

\clearpage

\end{document}